\documentclass[manuscript]{acmart}

\usepackage{multirow}
\usepackage{tabularx}
\usepackage{cleveref}
\usepackage{makecell}
\AtBeginDocument{%
  }

\setcopyright{acmlicensed}
\copyrightyear{2018}
\acmYear{2018}
\acmDOI{XXXXXXX.XXXXXXX}
\acmConference[ISS '26]{Make sure to enter the correct
  conference title from your rights confirmation email}{June 03--05,
  2018}{Woodstock, NY}
\acmISBN{978-1-4503-XXXX-X/2018/06}

\begin{document}

% \title{An Empirical Study of User Intent Disambiguation in LLM-Assisted Parameter-Driven Geometry Editing in Virtual Reality}
\title{Beyond Conversations: Spatially-Anchored Previews for Intent Disambiguation in LLM-Assisted Geometry Editing in Virtual Reality}
% \title{Stabilizing the Interaction Trajectory: An Empirical Study on Visual Scaffolding for Intent Disambiguation in Spatial Editing Workflows}

\author{Junlong Chen}
\orcid{1234-5678-9012}
\affiliation{%
  \institution{University of Cambridge}
  \city{Cambridge}
  \country{United Kingdom}
}
\email{jc2375@cam.ac.uk}

\author{Amr Gomaa}
\affiliation{%
  \institution{German Research Center for Artificial Intelligence}
  % \city{Hekla}
  \country{Germany}}
\email{Amr\_Gomaa\_Mohamed\_Elhady.Gomaa@dfki.de}

\author{Jens Grubert}
\affiliation{%
  \institution{Coburg University of Applied Sciences and Arts}
  \city{Coburg}
  \country{Germany}}
  \email{Jens.Grubert@hs-coburg.de}

\author{Per Ola Kristensson}
\affiliation{%
 \institution{University of Cambridge}
 \city{Cambridge}
 \country{United Kingdom}
 }
 \email{pok21@cam.ac.uk}

\renewcommand{\shortauthors}{Chen et al.}

\begin{abstract}
  User intent disambiguation remains a key challenge in intelligent interactive systems. While they have been widely studied in dialogue systems in 2D interfaces, research on how intent disambiguation could be incorporated within Large Language Model (LLM) assisted editing workflows in immersive environments remains limited. Recent advances in LLMs create opportunities to leverage the immersive nature of virtual and augmented reality (VR/AR) environments to provide better disambiguation support. In this paper, we evaluate how traditional dialogue-based disambiguation can be augmented with spatially-anchored graphical previews to resolve ambiguous user commands in LLM-assisted parameter-driven editing workflows. A within-subjects study in which 24 participants completed complex geometry editing tasks in VR simulate scenarios where VR scenes are controlled by numerical parameters. Compared with the condition where disambiguation is not available, quantitative metrics and qualitative feedback indicate that a hybrid approach which combines clarification questions and graphical previews can support better interaction stability with fewer conversation rounds while improving user experience. These findings provide empirical evidence on the effectiveness of disambiguation methods in LLM-assisted editing of parameter-driven immersive scenes and inform design guidelines for future integration of LLMs in advanced VR/AR systems.
\end{abstract}




\begin{teaserfigure}
  \includegraphics[width=0.33\linewidth]{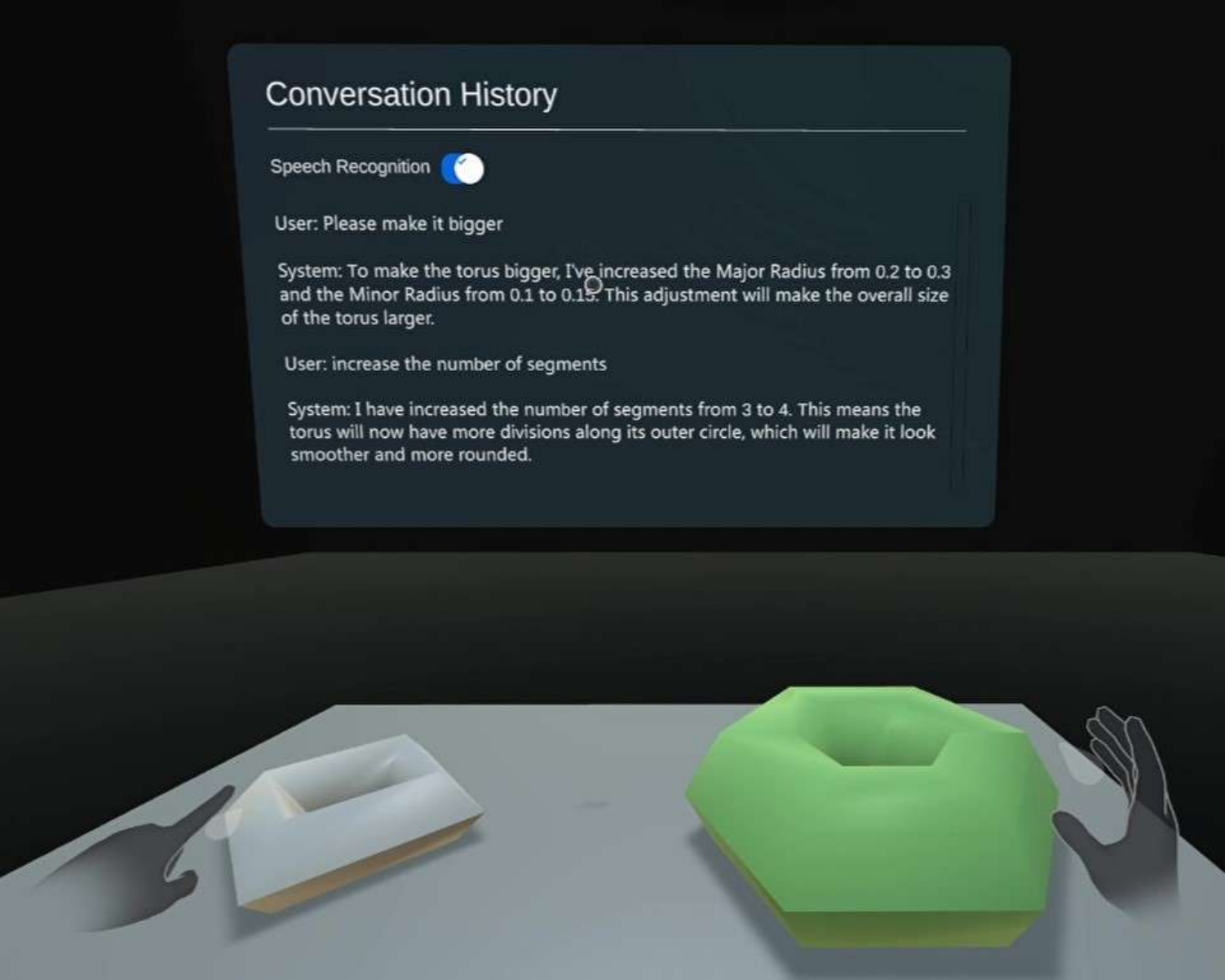}
  \includegraphics[width=0.33\linewidth]{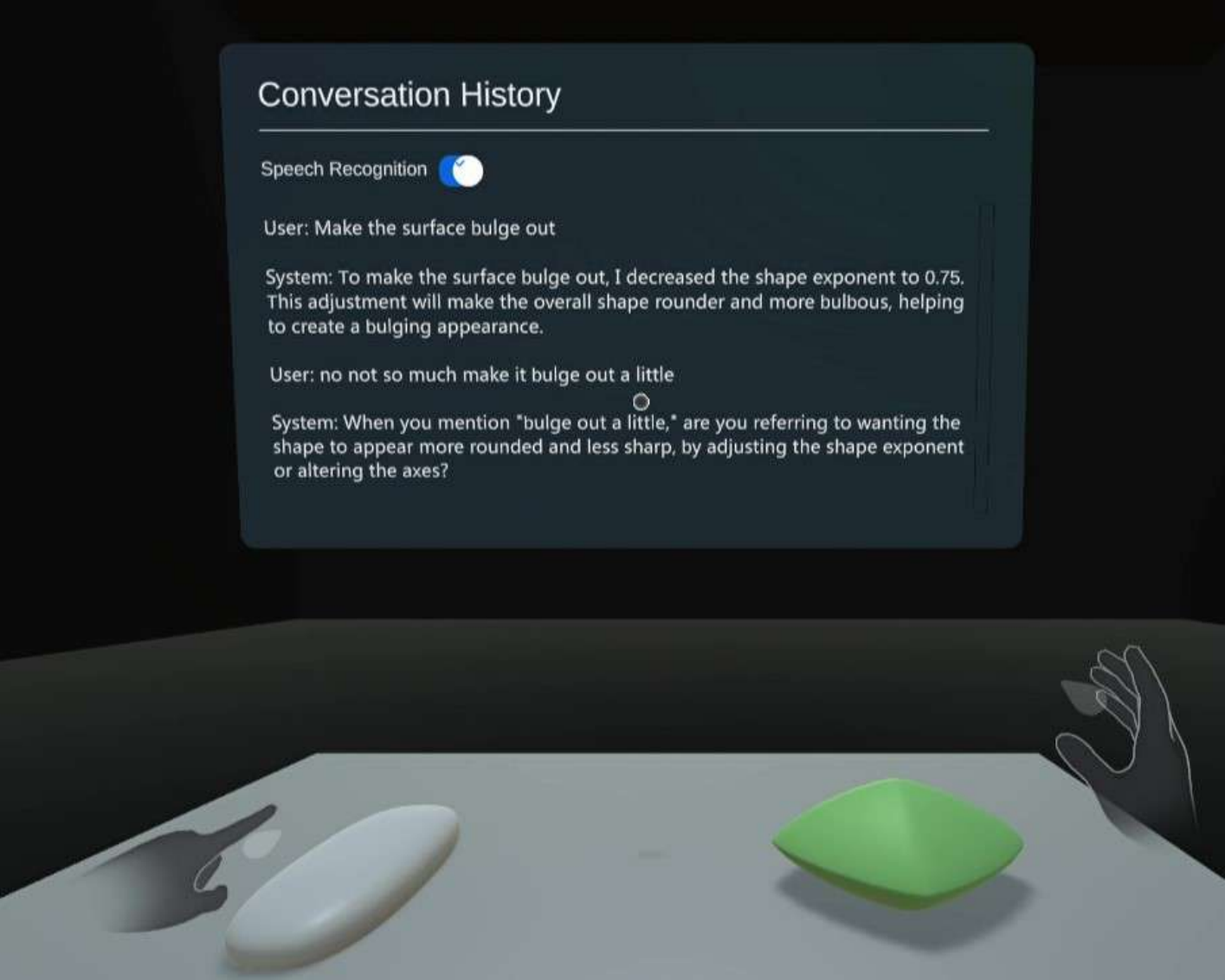}
  \includegraphics[width=0.33\linewidth]{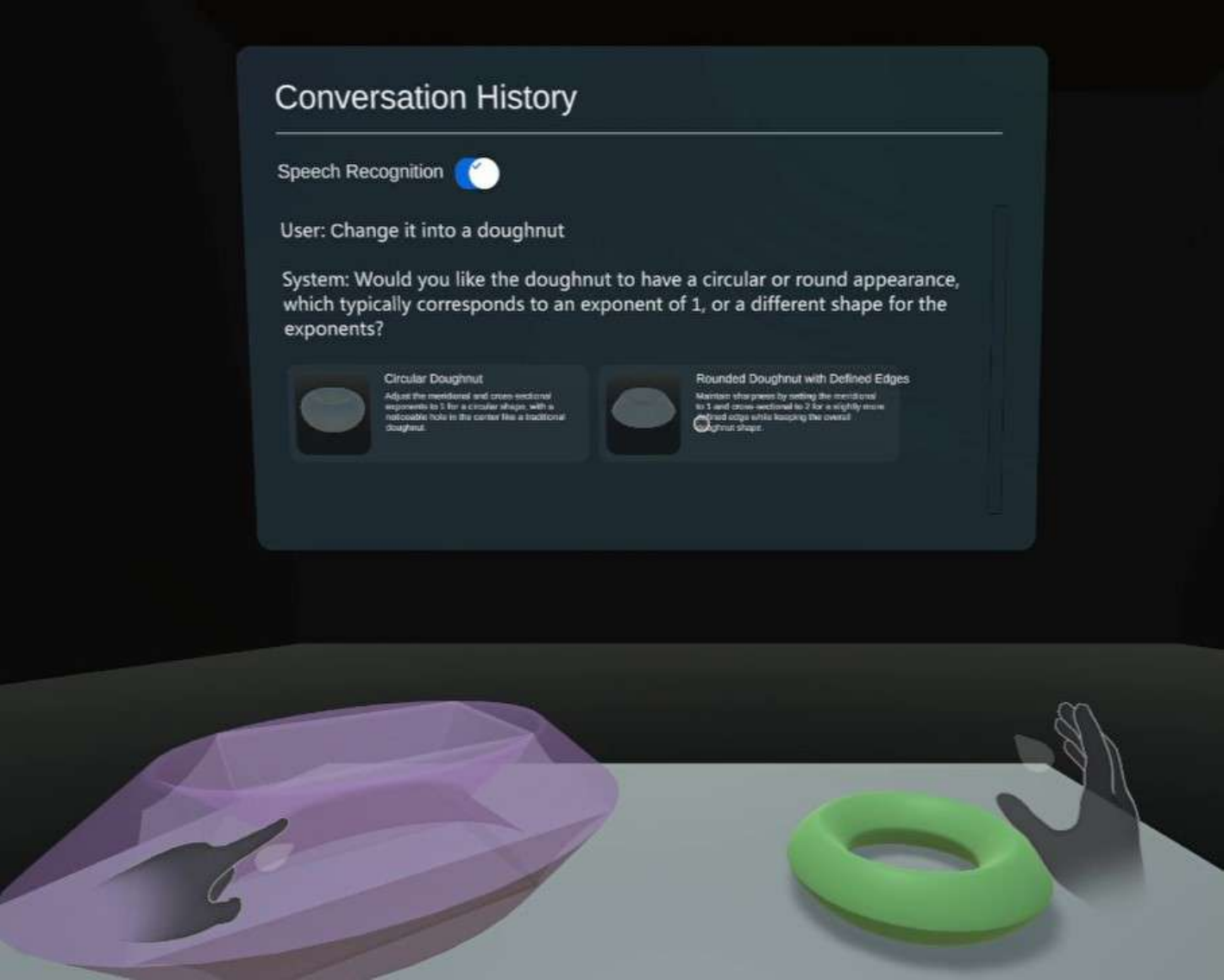}
  \caption{Examples of a user editing a polyhedral torus without disambiguation support (left), the user editing a superellipsoid with the system providing clarification questions (middle), and the user editing a supertoroid with the system providing clarification questions and graphical previews (right).}
  \Description{Examples of a user editing a polyhedral torus without disambiguation support (left), the user editing a superellipsoid with the system providing clarification questions (middle), and the user editing a supertoroid with the system providing clarification questions and graphical previews (right).}
  \label{fig:teaser}
\end{teaserfigure}


\maketitle

\section{Introduction}
Disambiguating user intent is crucial in interactive systems, especially in complex 3D spatial domains. Recent works have demonstrated the possibility of incorporating large language models (LLMs) in 3D environments and the potential benefits~\cite{chen2025analyzing, de2024llmr, giunchi2024dreamcodevr, tang2025llm, wang2024virtuwander}. However, an unaddressed challenge lies in limited user agency due to ambiguous user input and the lack of mechanisms integrated in LLM-assisted scene editing workflows to improve user agency and reduce errors~\cite{chen2025analyzing, zhang2024vrcopilot}. Various non-LLM systems have explored effective disambiguation through clarification questions \cite{alfieri2022intent, dhole2020resolving} and graphical previews \cite{li2020multi}. There is a missed opportunity to explore how these approaches could be adapted for spatial environments. In this work, graphical previews refer to a hybrid of 2D UI image previews and spatially-anchored 3D overlays. Unlike many speech-and-pointing works~\cite{bolt1980put, wong2010you, chen2025comparative, lee2024gazepointar} which study referential disambiguation, we clarify that this work does not study disambiguation on the input side of interaction. Instead, this work studies whether and how spatially-anchored feedback provide richer disambiguation support than what is possible in 2D interfaces, as well as its unique opportunities and challenges.

We explore how disambiguation techniques could be implemented within the context of LLMs for extended reality (XR), where the LLM controls a set of parameters which uniquely define an immersive scene. This idea follows existing parameterized shape editing workflows
% \footnote{Here, the geometry editing task is chosen to create a controlled environment for our study. This task serves as an example of parameter-driven editing tasks in VR/AR.}
~\cite{parsel}.
% and serves as an example of a wide variety of parameter-driven editing tasks that could be implemented in VR/AR. More prospective applications are discussed in \Cref{sec:discussion}.
To support our study, we designed \textsc{DisambVR}, a system that detects ambiguity in user speech commands and offers (a) clarification questions, and (b) 2D and 3D graphical previews. We adopt this system as a testbed to study disambiguation techniques in parameter-driven 3D editing.
% This work makes a conscious decision to focus on internal validity over external validity. As such, due to paper length restrictions, whether findings generalize to other parameter-driven editing tasks is not tested in this work and remains  as an open question.
% is discussed as future work in \Cref{sec:discussion}.
% We discuss in \Cref{sec:discussion} that this workflow might extend to more complex parameter-driven editing workflows in spatial environments as LLM capabilities improve.
% for LLMs to control 3D scenes is feasible and reasonably conjecture that as LLM capabilities improve, they will be able to process larger amounts of data to perform complex tasks in 3D, which will allow our findings to generalize to more complex scenarios. 
In our current study, the disambiguation techniques are evaluated in three 3D geometry editing tasks, where participants are invited to edit a 3D object to match the target shape of another 3D object of the same geometry type through speech commands. In these tasks, the user is not told that the geometry is defined by a set of parameters, nor are users briefed about the specific names and meanings of the parameters. This is an example of the need for systems to disambiguate the user's intent which is often unclear as parameter names are unknown. Here, the geometry editing task was deliberately chosen and simplified to include only four editable parameters. This is a conscious design decision to preserve internal validity, but this could in theory be expanded into more parameters which define more features in the immersive VR space.

To understand the evolution of LLM assistance in 3D environments, we adopted an additive study design. We included a baseline system without disambiguation support (\textsc{NONE} condition), stepped up to a traditional dialogue-based disambiguation system (\textsc{CQ} condition), and finally evaluated a multimodal system where dialogue is augmented with previews on a 2D UI and in-situ spatial graphical previews (\textsc{CQGP} condition). We conducted a within-subjects study with 24 participants and compared the three disambiguation conditions. More specifically, we studied the following research questions (RQs):

\begin{itemize}
    % \item \textbf{RQ1:} How does combining clarification questions and visual previews (\textsc{CQVP}) affect variability in task progression and interaction efficiency compared with clarification questions (\textsc{CQ}) alone or no disambiguation support (\textsc{NONE})?
    \item \textbf{RQ1:} Does adopting text-based clarification (CQ) affect variability in task progression, and does augmenting it with graphical previews (CQGP) further improve interaction efficiency?
    % \item \textbf{RQ2:} Compared with \textsc{CQ} and \textsc{NONE}, does \textsc{CQVP} reduce performance variability in task progression and the number of conversation rounds needed to complete the task?
    \item \textbf{RQ2:} How do clarification questions and graphical previews affect users' perceived workload, user experience, and sense of agency in editing tasks where ambiguity is present?
\end{itemize}

Our results show that the hybrid approach significantly improved user experience compared with the \textsc{CQ} condition. While user peak performance was not significantly different across the three different conditions, results showed a steadier improvement trajectory and significantly fewer conversation rounds under the \textsc{CQGP} condition. Qualitative results from post-experience questionnaires further reveal how \textsc{DisambVR} affects cognitive load and user agency, providing insight on why the hybrid approach outperformed clarification questions alone. These findings align with existing literature \cite{hu2025vision, oviatt2003user, zhang2009user} highlighting the need for adaptive, user-centered, and context-aware systems that flexibly respond to user needs. 

In summary, we make the following contributions: 

\begin{itemize}
    \item \textbf{C1:} We implemented \textsc{DisambVR} to support a controlled study of disambiguation techniques in LLM-assisted parameter-driven editing workflows in virtual reality. Through a within-subjects study, we found that the condition with full support (\textsc{CQGP}) resulted in a smoother task progression (significantly lower performance variability) compared with the \textsc{NONE} condition, significantly fewer conversation rounds compared with the \textsc{NONE} and \textsc{CQ} conditions, as well as higher user experience ratings compared with the \textsc{CQ} condition.
    \item \textbf{C2:} Based on the study results, we propose design recommendations specifically for user intent disambiguation in immersive LLM-assisted scene editing workflows, guiding developers seeking to integrate LLM-assisted disambiguation features in VR/AR platforms.
\end{itemize}

\section{Related Work}\label{sec:rel-work}

Our work builds upon previous works on traditional user intent disambiguation techniques in dialogue systems and 2D user interfaces (UIs) and intent disambiguation based on LLMs to explore how such disambiguation techniques could incorporate LLMs to provide user intent disambiguation in immersive 3D spaces.

\subsection{User intent disambiguation in dialogue systems}

Intent detection plays an important role in object manipulation and task oriented dialogue systems~\cite{arora2024intent}. Several approaches have been proposed for seamless intent detection and disambiguation for better user experience and enhanced system performance. The most widely-used disambiguation approaches are asking clarification questions~\cite{alfieri2022intent,zou2017understanding,farshidi2024understanding,dhole2020resolving,hu2020interactive,gan2020modeling,chi2024clarinet}. Alfieri et al.~\cite{alfieri2022intent} investigated the triggering point for asking clarification questions based on system uncertainty. Dhole~\cite{dhole2020resolving} proposed a rule-based system to generate discriminative questions for resolving ambiguities between intents in task-oriented dialogue systems. Zhang et al.~\cite{zhang2016mining} demonstrated the use of neural networks for refining user intents in multi-intent scenarios. Lastly, Zamani et al.~\cite{zamani2020generating} compared rule-based, neural network-based, and reinforcement learning-based methods to generate clarification questions for disambiguation in conversational systems. While these approaches focus on maximizing information gain to resolve ambiguities in user queries, they are restricted to traditional conversational methods and simple 2D interfaces. Additionally, they are not adaptable to changes in user behavior and learning effects for expert users who are already familiar with the system. Therefore, in this work, we expand on these techniques for 3D environments and study additional disambiguation techniques such as graphical previews~\cite{li2020multi} to assess user experience and performance.

\subsection{LLM-based user intent disambiguation}

With the current proliferation of LLMs, intent detection has become more accurate. However, LLMs are still highly susceptible to misspecification and require several trials to reach the desired output or action~\cite{tian2025fixing}. Therefore, disambiguation techniques are required to augment LLMs to perform the desired tasks swiftly and accurately. Several works have proposed augmenting off-the-shelf LLMs with prompt engineering and fine-tuning for intent disambiguation. Chi et al.~\cite{chi2024clarinet} introduced CLARINET, a novel framework for generating clarification questions in information retrieval systems by conditioning on retrieval distributions and fine-tuning large language models. This method outperformed traditional rule-based and machine learning-based approaches in retrieval success. 
Ning et al.~\cite{ning2024user} and Baek et al.~\cite{baek2024knowledge} augmented LLMs with user history to overcome the need for intent disambiguation. However, these methods are static (i.e., only general user demographics are considered) and are not adaptable to changing user behavior and current state of mind. Zhang et al.~\cite{zhang2023clarify} utilized LLMs to determine when to ask clarification questions, what kind of questions to ask, and how the LLM can respond accurately afterwards for a traditional conversational system. 
% Abugurain et al.~\cite{abugurain2024integrating} used the same approach for motion planning in the robotics domain.
Such systems, however, remain limited to the tested domains (due to fine-tuning) and they are solely usable for text-only dialogue systems. Additionally, they also focus on only generating clarification questions as the disambiguation technique. In our work, we perform an empirical study to investigate the effectiveness of clarification questions and graphical previews in user intent disambiguation in virtual reality. We demonstrate the feasibility of incorporating LLMs to provide 3D context awareness and adaptive support for complex editing tasks through a parameter-driven workflow.

\subsection{User intent disambiguation in 3D spaces}

While most of the previous approaches focus on user intent disambiguation for text-only dialogue systems and 2D interfaces~\cite{bolt1980put}, additional work has investigated multimodal intent disambiguation for 3D spaces, especially with AR/VR systems. 
graphical previews using either 2D images~\cite{chen2023arid, fu2022easyvrmodeling, vachha2025dreamcrafter}, miniature 3D models~\cite{mendes2017mid, zhang2023vrgit}, or semi-transparent 3D overlays~\cite{mathis2024mr} are commonly used for disambiguating different options in AR/VR menus.
Chen et al.~\cite{chen2020disambiguation} investigated combining other modalities (e.g., head, gaze, foot tap) with speech for intent disambiguation when interacting with several objects in VR.
Similarly, Liang et al.~\cite{liang2025handproxy} suggested using a virtual hand for intent disambiguation by expanding the supported affordances of a given dialogue system. Karli and Fitzgerald~\cite{karli2023extended} employed active learning and template-based methods to resolve ambiguities in LLM-enabled human-robot collaboration using clarifying questions and interpretation techniques to disambiguate user instructions. On the other hand, Chang et al.~\cite{chang20243d} utilized Multimodal Large Language Models (MLLMs) for improved 3D spatial understanding and contextual object localization and disambiguation in complex environments. 
While these approaches work for expert users with knowledge of the system design and parameters, they are not suitable for novice users who are neither familiar with the given system nor aware of the underlying design parameters. In this work, we focus on a single input modality approach (i.e., speech only) while utilizing a multiple output modality system (i.e., text and graphical previews) to resolve any misspecification and ambiguity regarding the given system and the underlying parameters.

\begin{figure*}[h!]
    \centering
    \includegraphics[width=\linewidth]{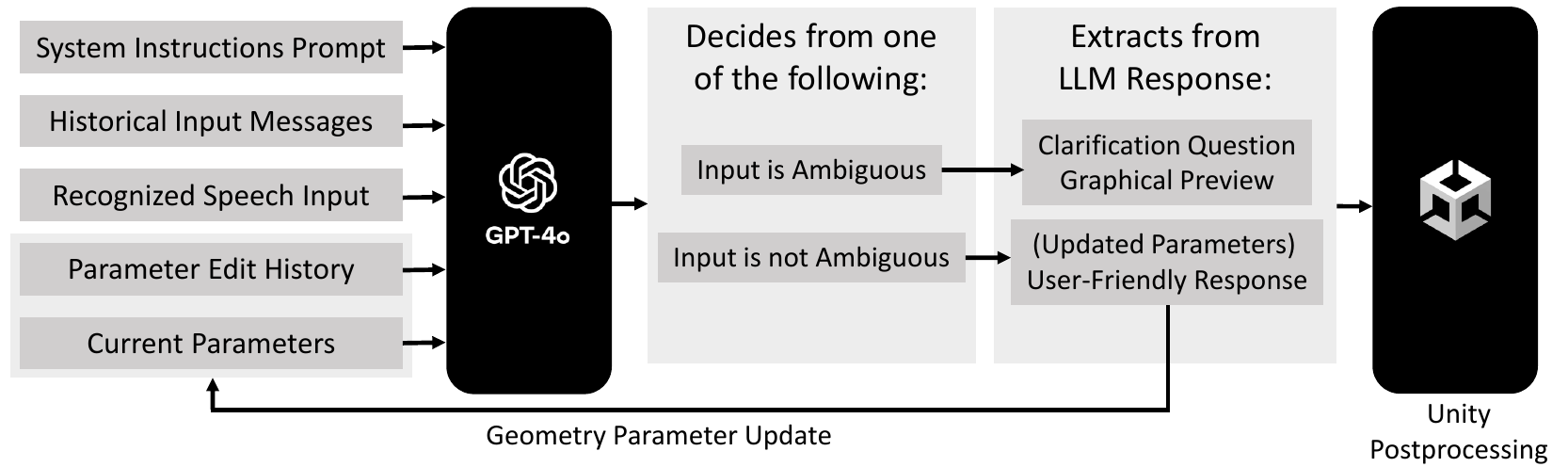}
    \caption{Workflow of \textsc{DisambVR}. The system feeds the system instructions prompt, historical input messages, the currently-recognized speech input, parameter edit history, and current parameters to GPT-4o. If the LLM decides that the input is ambiguous and can only be resolved with a low confidence, the clarification question and information for the graphical preview options are extracted. Otherwise if the input is not ambiguous, the updated parameters for the geometry (if an edit is made) and a user-friendly response are extracted. All extracted textual information are sent to postprocessing scripts in Unity to make updates to the geometry and/or display feedback information in the user interface.}
    \label{fig:workflow}
    \vspace{-1em}
\end{figure*}

\section{Method}\label{sec:methodology}

This section presents the implementation details of \textsc{DisambVR}, an LLM-assisted user intent disambiguation prototype built to solely provide different disambiguation techniques to support our study. 
Prior work \cite{mendes2017mid} has shown that previews alone can help, while dialogue systems \cite{tang2025llm} excel at natural language intent interpretation. We clarify that this work does not aim to compare all disambiguation paradigms, but rather focuses on whether combining clarification questions and graphical previews outperforms clarification questions alone in an immersive geometry editing context. This section will highlight how this system builds upon recent advances in LLM interfaces \cite{Google_2023, Pichai_2024} as well as user intent disambiguation techniques applied in 2D interfaces \cite{alfieri2022intent, li2020multi}, but differs from traditional 2D UI by providing additional feedback through a combination of 2D and 3D graphical previews.

Through a user study where participants edit the appearance of complex 3D geometries in VR through speech commands, we gather log data to calculate performance metrics and use feedback from questionnaires to assess perceived task load and user experience. The study is approved by the research ethics committee at $<$\textit{anonymized for review}$>$.

\subsection{Apparatus}

To study the effects of disambiguation techniques in 3D environments, we designed \textsc{DisambVR}. The system aims to resolve ambiguous commands in VR through two primary disambiguation techniques:

\begin{enumerate}
    \item \textbf{Clarification Questions (CQs):} A short query such as “Do you mean the major radius or the minor radius?” appears on the UI in front of the user. Users respond verbally to resolve the ambiguity.
    \item \textbf{Graphical Previews (GPs):} A few options are provided in the conversation UI with a 2D image preview of the edited 3D geometry, together with a short heading and brief description to introduce the edit. When users hover over the option, a 3D preview of the edit is displayed in situ on the original object.
\end{enumerate}
These techniques are integrated within the Unity workflow, with the overall system workflow shown in \Cref{fig:workflow}. The LLM-assisted parameter-driven editing workflow is similar to existing `JSON-in, JSON-out' workflows~\cite{chen2025llmer, zhang2024vrcopilot, yang2025llplace} which represent the scene as structured data and pass it to the LLM, before receiving output from the LLM and converting the updated structured data to the spatial edit and outputting information for the user.

In \textsc{DisambVR}, speech input is recognized and sent to GPT-4o together with a textual system instructions prompt, historical input messages, the parameter edit history, as well as the current parameters of the geometry. The system prompt contains: (i) basic information on the geometry and its parameters, as well as (ii) instructions on the textual output format. 

For the former prompt (i), the LLM is provided with the context of the task, including the type of the geometry, the type of parameters which uniquely define this geometry and the range of the parameters, the latest user input, the latest parameter values for the geometry, as well as the edit history of all parameters. More details about the task will be provided in \Cref{sec:task}.

For the latter prompt (ii), the LLM is instructed to identify whether the user input is ambiguous or not ambiguous. If the input is ambiguous and can only be resolved with a low confidence, the LLM will output a textual response with a clarification question and a few options for the geometry parameters with a short heading and a brief description for each option. Otherwise, if the input is not ambiguous, the LLM will output the updated parameters (if any) and a user-friendly response. The LLM is instructed to format the response, which facilitates Unity post-processing scripts to handle the textual response and update the geometry based on the new parameters or provide feedback information (text or graphical preview options) in the UI in front of the user. 
To mitigate LLM hallucinations, our system parses only structured response segments using strict markers. Parameters are applied only if all four parameters are extracted properly, belong to the correct data type (integer/float) and fall within the correct range. \textsc{CQ}s and \textsc{GP}s also intend to provide users with more agency over the potentially-hallucinated edit options.
More details on the system prompt are provided in the Supplemental Materials. 

\begin{figure}[h!]
    \centering
    \includegraphics[width=0.6\linewidth]{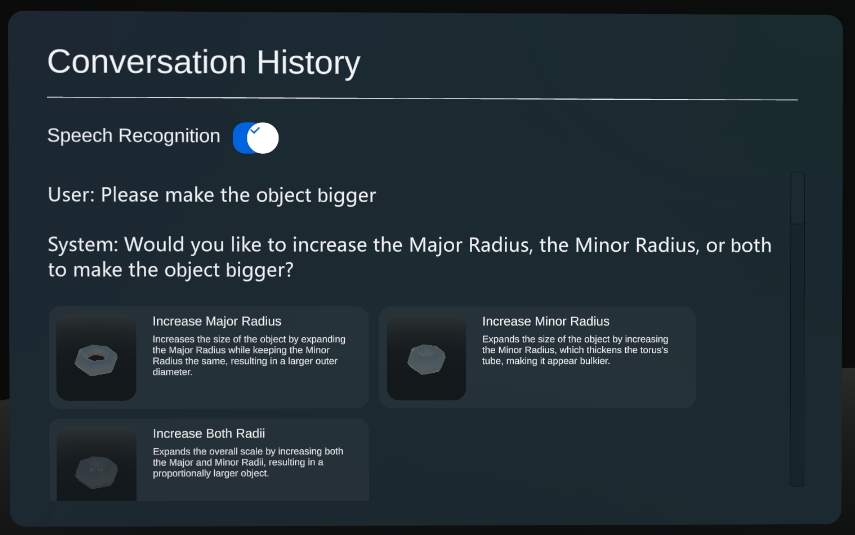}
    \includegraphics[width=0.6\linewidth]{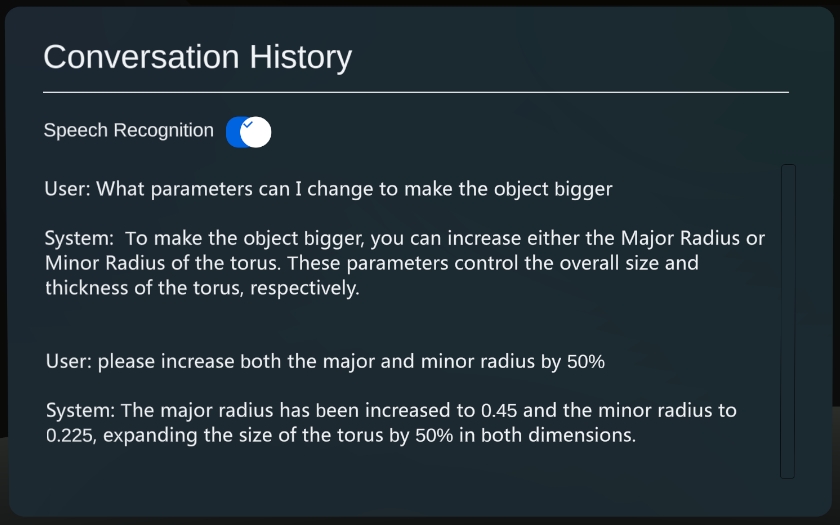}
    \caption{An example of the \textsc{DisambVR} system UI. The user toggles the speech recognition button on and asks the system to make the object bigger. The system detects ambiguity in the input and asks the user to clarify which parameter to increase, followed by three options with graphical previews, a heading, and a brief description of the edit to be made. The user is not satisfied with either of the options, and asks how can the geometry be made bigger. The system provides a response to the user query. Next, the user asks the system to increase the major and minor radius by 50\%. The system does not detect ambiguity in the input so it makes the edit directly and outputs a message to describe the action taken.}
    \label{fig:ui}
    \vspace{-1.5em}
\end{figure}

This system is implemented based on the Meta Interaction SDK \cite{Meta_2025} and uses the built-in gestures to control the UI in front of the user. The UI is also developed based on a UI template in the Meta Interaction SDK package. An example of the UI is provided in \Cref{fig:ui}. 
The user interface panel is located in front the user, and consists of the following elements:
\begin{enumerate}
    \item \textbf{Speech Recognition Toggle:} When enabled, the system will perform continuous speech recognition. There is no additional button to indicate the start or end of speech recognition. 
    % It is switched off by default, and users can toggle the button on when they are ready after they enter the scene.
    \item \textbf{User Input:} Displays the recognized natural language input in real time as the user speaks. The UI with input and output messages can be scrolled to view the full conversation history.
    \item \textbf{System Output:} If the user input is ambiguous, the system provides a clarification question accompanied by a few graphical preview options underneath it. The options consist of a 2D preview image in the left, and a heading and brief description in the right. When the user hovers the cursor over the option, a 3D preview of the edited geometry will be displayed in purple. If the user input is not ambiguous, the UI displays a message to explain the edit in natural language.
\end{enumerate}

Participants wore an Oculus Quest 2 headset and used  gestures to control the UI. The headset was connected to a Windows 10 laptop PC (Intel i5-9300H CPU, 16GB memory, and GTX 1050 graphics card) using an Oculus link cable. The scene was implemented using Unity 3D (Version 2022.3.15f1).

\subsection{Participants}

24 participants (15 male, 8 female, and 1 preferred not to disclose) were recruited in the study. Their age ranged from 19 to 42 ($M=25.4, SD=5.9$). Around 20.8\% of participants reported being experienced or very experienced with head-mounted virtual reality, 20.8\% reported being experienced with speech recognition systems, and 37.5\% reported being experienced or very experienced with 3D user interfaces or similar immersive systems like CAD, video games, and simulation software. All participants understood and spoke English, with 37.5\% reported being native English speakers. None of the participants reported any form of disability which would affect their participation in the study. All participants provided informed consent and were remunerated after the study in appreciation of their participation.

\subsection{Task}\label{sec:task}

The study adopted a within-subjects design where each of the 24 participants were invited to edit three different types of 3D geometries to match a target appearance. \Cref{fig:teaser} shows the setup of the task from the user's field-of-view, including all three conditions and all three types of geometries tested in the study.
Multiple geometry types allow us to test whether users can perform similarly across different geometries.
% \begin{figure}[h!]
%     \centering
%     \includegraphics[width=\linewidth]{figures/task-setup.png}
%     \caption{User field-of-view of the task setup. The white original geometry and green target geometry are placed on a desk in front of the user. In front of the user behind the geometries is the UI panel which provides the toggle button for continuous speech recognition and displays the conversation history.}
%     \label{fig:scene-setup}
% \end{figure}
Here, the white geometry on the left is the original geometry, and the green geometry on the right is the target appearance. Participants are instructed only to edit the shape and are asked not to change the color of the geometry. The initial viewpoint of all users at the beginning of each trial was fixed to preserve internal validity, and participants were allowed to adjust their posture to view the geometry at slightly different angles.
The geometries involved in the tasks, together with descriptions of the parameters which control each geometry are provided in \Cref{tab:geometry}. For consistency, all geometries are uniquely determined by four parameters, with the first two being continuous parameters with the same range and the latter two being discrete parameters with the same number of available values to take.

\begin{table}[h!]
  \caption{Geometries and parameters involved in the tasks.}
  \label{tab:geometry}
  \renewcommand{\arraystretch}{1.5}
  \resizebox{\columnwidth}{!}{
\begin{tabularx}{1.25\columnwidth}{p{0.35\columnwidth}X}
\hline
\textbf{Geometry/Parameter}          & \textbf{Description}                                                                                          \\ \hline
\textbf{Polyhedral   Torus}          & \textbf{A three-dimensional polygonal   approximation of a torus.}                                             \\ \hline
Major   Radius ($R$)                 & Distance from the torus center   to the center of the tube, range 0.1 to 0.5.                                  \\
Minor   Radius ($r$)                 & Radius of the tube itself, range   0.1 to 0.5.                                                                 \\
Segments   ($N$)                     & Number of divisions along the   major circular path, integer range 3 to 10.                                    \\
Radial   Segments ($M$)              & Number of divisions along the   tube's cross section, integer range 3 to 10.                                   \\ \hline
\textbf{Superellipsoid} &
  \textbf{Scaling a ``basic shape" along three perpendicular axes and   ``rounding" it using a shape exponent.} \\ \hline
Semi-Major Axis ($a$)                & Center to surface along the   left-to-right direction, range 0.1 to 0.5.                \\
Semi-Minor   Axis ($b$)              & Center to surface along the bottom-to-top direction, range 0.1 to 0.5.                \\
Semi-Intermediate Axis ($c$) &
  Center to surface along the front-to-back direction. $c \in \{0.15, 0.2, 0.25, 0.3,   0.35, 0.4, 0.45, 0.5\}$ \\
Shape Exponent ($\epsilon$)          & Controls the ``squareness" or ``roundness" of the   shape. $\epsilon\in \{0.25, 0.5, 0.75, 1, 1.5, 2, 2.5, 3\}$ \\ \hline
\textbf{Supertoroid} &
  \textbf{Doughnut-like surfaces which can be generalized by ``super-shaping" its cross section and meridional profile.} \\ \hline
Major Radius ($R$)                   & Distance from the center of the torus to the center of the   tube, range 0.1 to 0.5.                          \\
Minor Radius ($r$)                   & Nominal radius of the tube, range 0.1 to 0.5.                                                                 \\
Meridional Exponent   ($\epsilon_1$) & Controls the shape in the meridional direction.   $\epsilon_1\in\{0.25, 0.5, 0.75, 1, 1.5, 2, 2.5, 3\}$         \\
Cross-Sectional Exponent   ($\epsilon_2$) &
  Controls the shape of the tube's cross section.   $\epsilon_2\in\{0.25, 0.5, 0.75, 1, 1.5, 2, 2.5, 3\}$ \\ \hline
\end{tabularx}
    }
\end{table}

\begin{figure}[h!]
    \centering
    \includegraphics[width=0.2\linewidth]{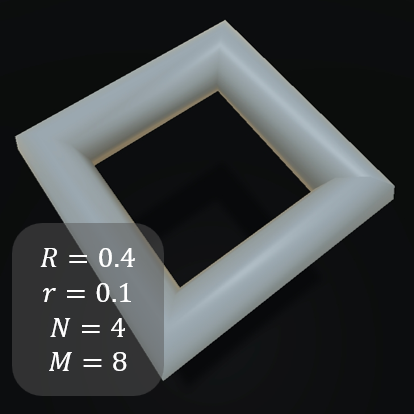}
    \includegraphics[width=0.2\linewidth]{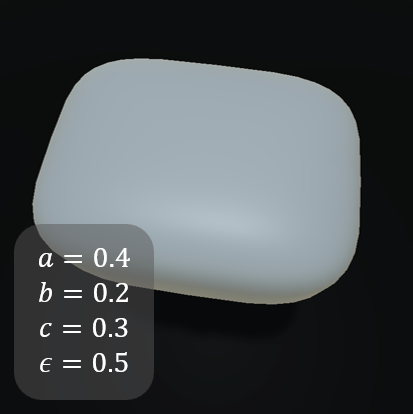}
    \includegraphics[width=0.2\linewidth]{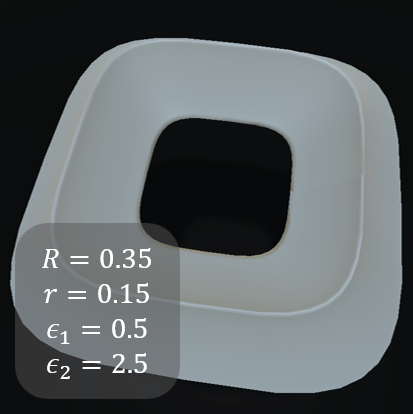}\\
    \includegraphics[width=0.2\linewidth]{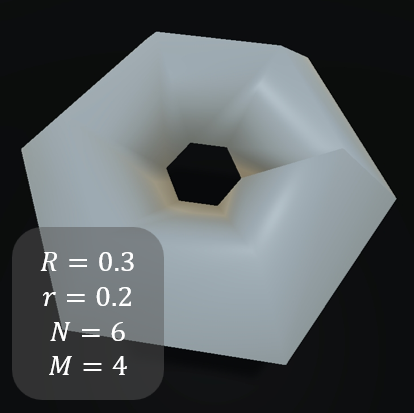}
    \includegraphics[width=0.2\linewidth]{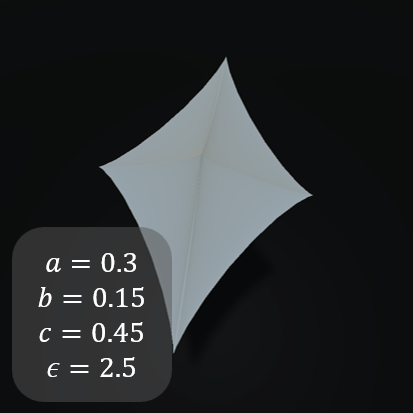}
    \includegraphics[width=0.2\linewidth]{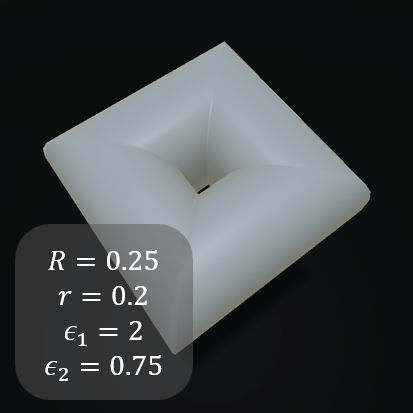}
    \caption{Examples of the polyhedral torus (left), superellipsoid (middle), and supertoroid (right). The values of the four parameters which define the geometry are listed in each subfigure.}
    \label{fig:geometries}
    \vspace{-1em}
\end{figure}

Prior to the study, participants were not informed of the types of geometries or any information about the parameters which control the geometries. This is intentionally designed to result in ambiguity in user commands. Examples of these geometries together with their parameters are shown in \Cref{fig:geometries}.

\subsection{Procedure}

At the beginning of the study, participants were greeted and invited to read an information sheet and sign a consent form. Participants filled out a brief demographics survey and received an introduction of the study. This involved explaining the tasks, number of conditions and trials in each condition, and a very basic introduction on how to use the system. Participants were only instructed on how to use gestures to control the UI and were advised that they could ask questions or issue commands directly using speech. Details on the types of geometries and how geometries can be determined by setting parameters were not shared with participants at the beginning of the study. Participants were encouraged to speak to the system to find out more about how to use it. Details on the instructions given to participants at the beginning of the study can be found in the Supplemental Materials.

Subsequently, participants were instructed how to put on and adjust the Oculus Quest 2 headset, and entered the scene to complete the geometry editing tasks under different conditions. A within-subjects design was adopted and each of the 24 participants completed all three conditions. The prompts for different conditions in \textsc{DisambVR} are provided in the Supplemental Materials. The three conditions are:
\begin{itemize}
    \item \textbf{\textsc{NONE} Condition:} The default condition without any disambiguation support. However, the system is still powered by an LLM. For all speech commands it receives, the system tries to make geometry edits directly, unless if the input is a query where the system will provide a response.
    \item \textbf{\textsc{CQ} Condition:} If the LLM assesses that the user input is very ambiguous, the system will output a clarification question (CQ) to ask the user to clarify their intent. If the user input is not ambiguous, the system will make edits to the geometry and provide a message to explain the change made. If the user input is a query, the system will provide a natural language response.
    \item \textbf{\textsc{CQGP} Condition:} In addition to the \textsc{CQ} condition, if the user input is very ambiguous, the system will provide a few additional graphical preview (GP) options below the system response text. Each option contains a 2D graphical preview on the left and a heading and brief description on the right. When users hover over the option, the original white geometry will become semi-transparent, and a 3D purple preview will be overlaid at the original object position. Users double-click on the option to confirm the edit, or they can choose not to make any edits and initiate a new round of conversation.
\end{itemize}

Within each condition, participants first spent no more than 1.5 minutes to familiarize with the system in a practice trial. Subsequently, they spent no more than 4 minutes on Trial One and no more than 4 minutes on Trial Two. If participants were satisfied with their edit, they could end the trial before reaching the time limit. As participants worked through the task, the system logged timestamped data including recognized speech, system response, and the parameter values each time an edit was made.
To minimize learning effects, a different type of geometry was used in each condition. Across different trials, the original geometry parameter values and target geometry parameter values were also different, but the difference in the original and target parameter values was approximately the same to facilitate comparison across different trials and conditions.
The sequence of the three conditions and the three geometries was counterbalanced across all 24 participants such that participants experienced different combinations of condition and geometry. The complete sequence for all participants is provided in the supplemental materials. After completing each condition, participants completed a short questionnaire to reflect on their editing strategy and share their comments on the technique provided in this condition. They also completed an unweighted NASA-TLX questionnaire~\cite{hart1988development} and a UEQ-S~\cite{schrepp2017design} survey.

After completing all three conditions, participants completed a post-experience questionnaire to share any differences they perceived when they interacted with the system with disambiguation techniques and the system without disambiguation techniques. Participants were invited to share their preference between the \textsc{CQ} and \textsc{GP} disambiguation techniques and were asked to share their feedback on the two techniques. After the study, participants were thanked for their time and remunerated. The entire session lasted approximately 60 to 75 minutes.

% \section{Study Design}\label{sec:study-design}

\section{Results}\label{sec:results}

In this section, we summarize the quantitative and qualitative findings from our user study. We report metrics related with task performance, perceived workload, user experience, as well as results from a thematic analysis on the subjective comments on different disambiguation techniques.

\subsection{Task Performance}\label{sec:performance}

Following the average parameter prediction errors (MAE) for normalized scalar parameters proposed by Hossain et al.~\cite{hossain2023data} as well as works on procedural 3D shape modeling~\cite{hossain2025approximating}, we propose a normalized parameter distance metric to evaluate the closeness between the current shape parameters and the target shape parameters in each trial. It provides a consistent and interpretable measure of editing accuracy across the three different types of geometries. For each parameter $p$, its normalized difference from the target is calculated as:

\begin{equation}
    \Delta{p}=\frac{|p_{current}-p_{target}|}{p_{max}-p_{min}},
\end{equation}

\noindent
where $p_{min}$ and $p_{max}$ are the minimum and maximum values allowed for parameter $p$. This yields a non-negative dimensionless distance metric for each parameter. Aggregating the normalized differences across all four parameters and subtracting it from one, we obtain a closeness score for the set of parameters at any point in time:

\begin{equation}
    \mbox{Closeness Score}=1-\frac{1}{4}\sum_{i=1}^4\Delta{p_i},
\end{equation}

\noindent
where $\Delta{p_i}$ is the normalized difference for parameter $i$. The closeness score reflects how close the current parameters are to the target parameters. It is a non-negative number and the closer it is to one, the better the user's performance is in matching the geometry parameters with the target. For each trial, we use the maximum closeness score as a metric to evaluate how close the participant was able to match the geometry with the target appearance. Participant scores are averaged across both trials.

Conventional paired $t$-tests revealed no statistically significant differences in the maximum closeness scores between the \textsc{NONE} and \textsc{CQ} conditions ($t_{(23)}=1.97$, $p=.061$), the \textsc{NONE} and \textsc{CQGP} conditions ($t_{(23)}=0.15$, $p=.880$), or the \textsc{CQ} and \textsc{CQGP} conditions ($t_{(23)}=-1.55$, $p=.134$). Friedman tests also did not reveal a significant difference across the three conditions ($\chi^2=4.75$, $p=.093$).

We further conducted two one sided t-tests (TOST)~\cite{lakens2018equivalence} to test for equivalence in the maximum closeness scores between the \textsc{NONE}, \textsc{CQ}, and \textsc{CQGP} conditions. Unlike difference tests, TOST requires an a priori equivalence bound to reflect the smallest effect size to be considered meaningful. We specified a single bound of $\pm0.03$ and applied this bound to all three pairwise comparisons. This bound was fixed by the research team prior to data collection as a reasonable benchmark convention and is not derived from pilot data or observed results~\cite{lakens2018equivalence}. Additionally, Holm correction was applied across the three TOST tests to control for multiple comparisons. Under this bound, none of the three pairwise comparisons revealed statistical equivalence after correction: \textsc{NONE} vs. \textsc{CQ} ($p_{\text{TOST,Holm}}=.873$), \textsc{NONE} vs. \textsc{CQGP} ($p_{\text{TOST,Holm}}=.124$), and \textsc{CQ} vs. \textsc{CQGP} ($p_{\text{TOST,Holm}}=.873$).

A post-hoc sensitivity analysis was conducted to interpret this null result. At sample size $n=24$, our design had around 8-24\% power (24.0\% power for \textsc{NONE} vs. \textsc{CQ}, 22.1\% power for \textsc{NONE} vs. \textsc{CQGP}, and 8.4\% \textsc{CQ} vs. \textsc{CQGP}) to detect equivalence within the $\pm0.03$ bound, which is well below the conventional 80\% threshold. Reaching an adequate power requires around 56 to 72 participants, and we therefore do not interpret these results as evidence of equivalence in peak performance across conditions.

Regarding geometry types, Friedman tests did not reveal a significant difference in maximum closeness scores ($\chi^2=1.33, p=.51$) across the three geometry types. Applying the same TOST procedure with Holm correction to pairwise geometries also failed to yield equivalence, and the similar sensitivity limitations above also apply. We further note that the condition and geometry are fully counterbalanced across the study, which rules out the possibility of a systematic condition-geometry confound in the comparisons above.

While equivalence tests were statistically underpowered, the lack of significant differences in peak performance across conditions suggests that given enough time, users may eventually brute-force their way to the target geometry regardless of the disambiguation condition. The true difference lies in the friction of that journey, which we highlight next.

\subsection{Interaction Data Analysis}\label{sec:interaction-data}

While according to the closeness score metric the three different disambiguation conditions did not yield a significant difference in final task performance, other observations can be made throughout the interaction process.

\begin{figure*}[h!]
    \centering
    \includegraphics[width=0.3\linewidth]{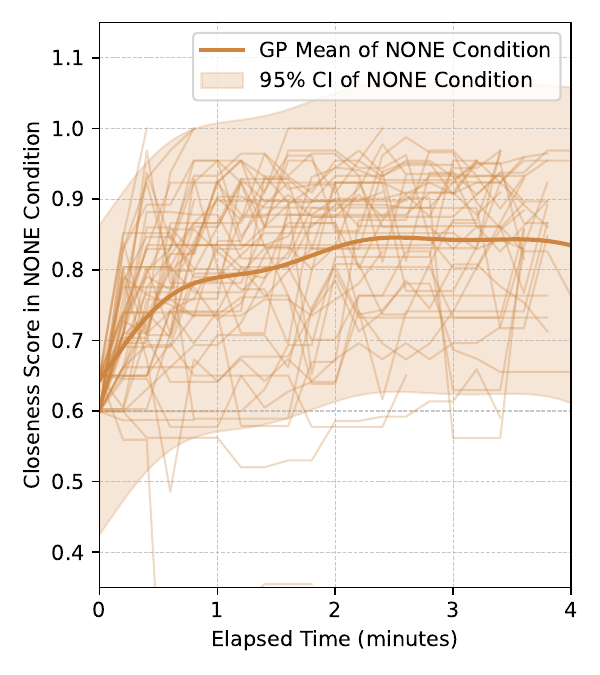}
    \includegraphics[width=0.3\linewidth]{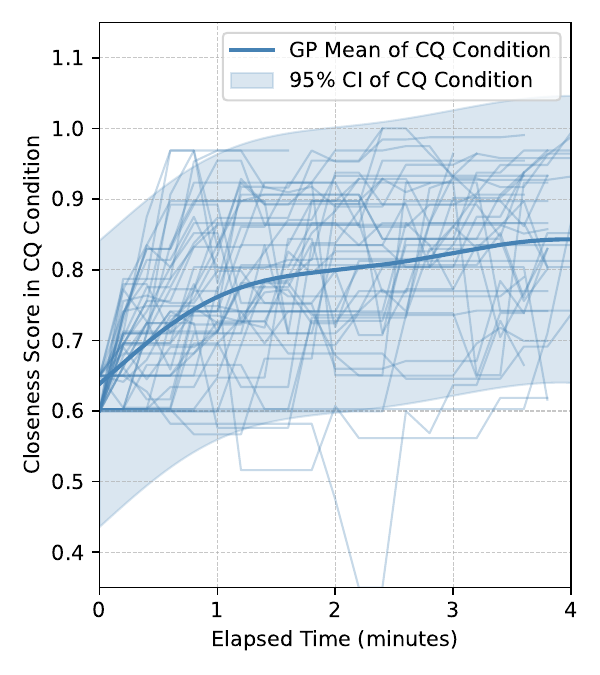}
    \includegraphics[width=0.3\linewidth]{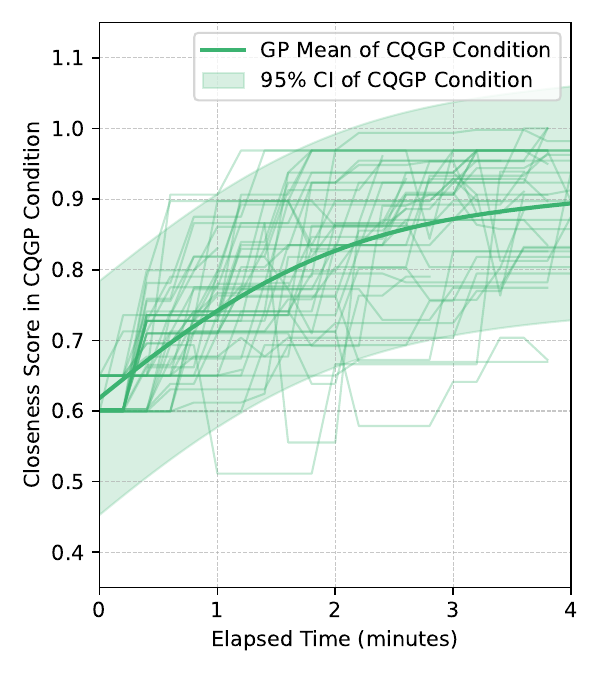}
    \caption{Progression lines showing the closeness score with respect to elapsed time in each trial conducted by each participant under the \textsc{NONE}, \textsc{CQ}, and \textsc{CQGP} conditions. A Gaussian Process (GP) Regression \cite{carl-gp} was applied to the elapsed time and closeness score data for each condition, with the
    % and a kernel combining a constant factor and a radial basis function was used to capture the trend, while a white noise kernel was used to account for noise. The Gaussian Process 
    predicted mean curve and 95\% confidence interval overlaid on the progression lines.}
    \label{fig:progression}
    \vspace{-1em}
\end{figure*}

\paragraph{Task Progression.} \Cref{fig:progression} plots the closeness score of all 24 participants over elapsed time for the \textsc{NONE}, \textsc{CQ}, and \textsc{CQGP} conditions. While participants completed the task using approximately the same amount of time (given in minutes) across the \textsc{NONE} ($M= 3.49, SD= 0.68$), \textsc{CQ} ($M= 3.63, SD= 0.68$), and \textsc{CQGP} ($M= 3.08, SD= 0.93$) conditions, 
the closeness score fluctuated more in the \textsc{NONE} ($M=5.24\times 10^{-3},SD=3.31\times 10^{-3}$) and \textsc{CQ}  ($M=3.18\times 10^{-3}, SD=1.92\times 10^{-3}$) conditions, as compared to the \textsc{CQGP} ($M=2.50\times 10^{-3}, SD=1.76\times 10^{-3}$) condition where participants made smoother task progression. 

Hence, the Mean Square of Successive Differences (MSSD) \cite{von1941mean} is used as a measurement of variability which considers the differences between consecutive values instead of just the overall spread. It is calculated as:

\begin{equation}
    \delta^2=\frac{\sum_{i=1}^{n-1} (x_{i+1}-x_i)^2}{n-1},
\end{equation}

\noindent
where $i$ refers to the temporal order of the closeness score $x_i$, which is recorded each time a change in closeness score occurs. $n$ is the total number of different closeness scores in one trial. The MSSD values are obtained at a same sampling rate of 5 Hz in the four-minute trials across the three conditions and averaged across the two trials resulting in a single MSSD value for each (Participant, Condition) combination.

\begin{figure*}[h!]
    \centering
    \includegraphics[width=0.25\linewidth]{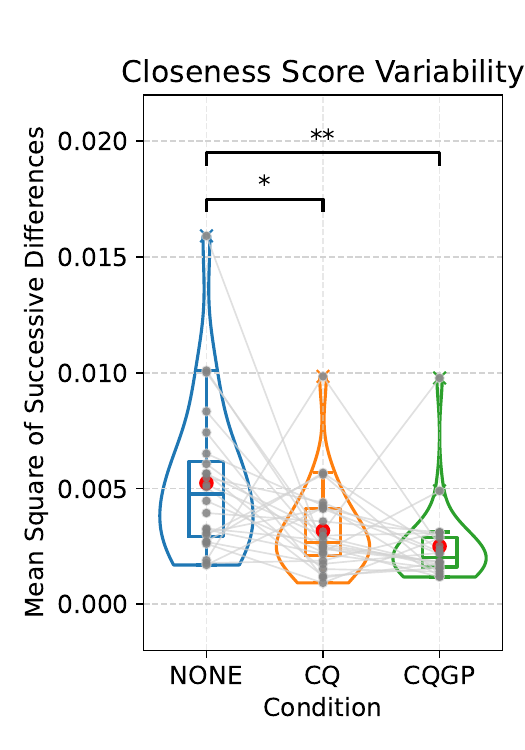}
    \includegraphics[width=0.25\linewidth]{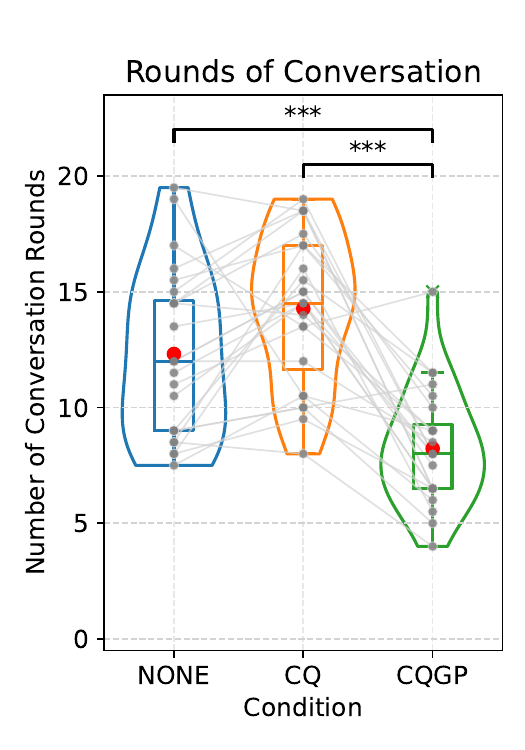}
    \includegraphics[width=0.25\linewidth]{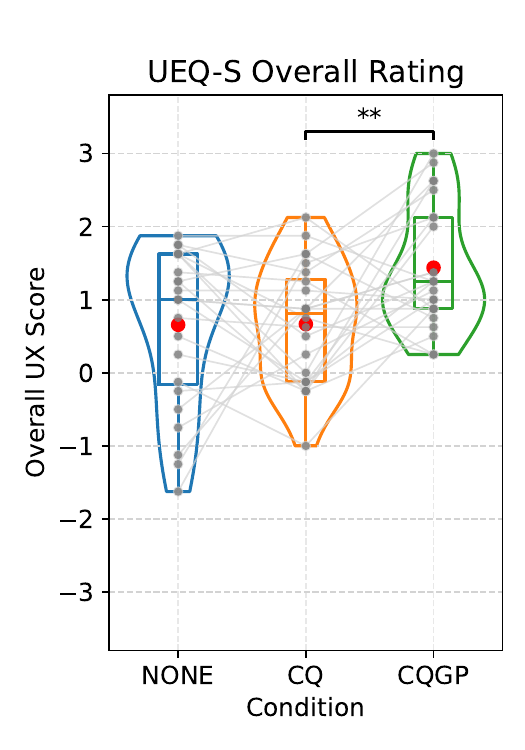}
    \caption{Violin plots for quantitative measures in the study where significant differences were found. Red dots indicate the arithmetic mean and asterisks indicate the level of significance in post-hoc tests: *$p<.05$, **$p<.01$, ***$p<.001$. Data from the same participant is connected with thin grey lines across the three conditions.}
    \label{fig:nasa-tlx-ueq-box}
    \vspace{-0.8em}
\end{figure*}

\textbf{Friedman tests revealed a significant difference ($\chi^2=10.3, p<.01$) in the MSSD values between the \textsc{NONE}, \textsc{CQ}, and \textsc{CQGP} conditions.}
Post-hoc Conover pairwise comparisons revealed that compared with the \textsc{NONE} condition, the MSSD values are significantly smaller in the
\textsc{CQ} ($T=2.23, p<.05$) and \textsc{CQGP} ($T=3.51, p<.01$) conditions, but the difference is not significant between the \textsc{CQ} and \textsc{CQGP} conditions ($T=1.28, p=.21$), as shown in \Cref{fig:nasa-tlx-ueq-box} (left).

\paragraph{Number of Conversation Rounds.} \textbf{Friedman tests indicated a significant difference ($\chi^2=25.2, p<.001$) in the rounds of conversations of the \textsc{NONE} ($M=12.3, SD=3.57$), \textsc{CQ} ($M=14.3, SD=3.23$), and \textsc{CQGP} ($M=8.23, SD=2.48$) conditions.} Post-hoc Conover comparisons revealed a significant difference in the number of conversations between the \textsc{NONE} and \textsc{CQGP} condition ($T=4.02, p<.001$), as well as the \textsc{CQ} and \textsc{CQGP} conditions ($T=7.11, p<.001$), as shown in \Cref{fig:nasa-tlx-ueq-box} (middle).

\subsection{Workload and User Experience}\label{sec:load-exp}

In terms of perceived workload, a Shapiro-Wilk test revealed that the \textsc{Overall} NASA-TLX score was normally distributed ($p>.05$). A repeated-measures ANOVA revealed no significant effect of \textsc{Condition} on overall NASA-TLX scores, ($F_{1.5, 35.0} = 1.29, p = .28, \eta_p^2 = .053$). Mauchly’s test indicated that the assumption of sphericity was violated ($W = 0.68, p = .02$), and Greenhouse-Geisser correction was applied here ($\varepsilon = 0.76$).

Regarding user experience, a Shapiro-Wilk test suggests non-normality ($p<.05$) in the overall UEQ-S ratings for the \textsc{NONE} condition. \textbf{Friedman tests were applied, which revealed a significant difference ($\chi^2=6.88, p<.05$) in the overall UEQ-S scores across the \textsc{NONE} ($M=.66, SD=1.07$), \textsc{CQ} ($M=.67, SD=.81$), and \textsc{CQGP} ($M=1.44, SD=.84$) conditions.} Post-hoc Conover pairwise comparisons revealed a significant difference ($T=2.74, p<.01$) in the overall UEQ-S score between the \textsc{CQ} and \textsc{CQGP} conditions.
Friedman tests also revealed a significant difference in the \textsc{Pragmatic} quality ($\chi^2=6.75, p<.05$) across the \textsc{NONE} ($M=.48, SD=1.04$), \textsc{CQ} ($M=.34, SD=1.09$), and \textsc{CQGP} ($M=1.34, SD=1.23$) conditions, as well as a significant difference in the \textsc{Hedonic} quality ($\chi^2=6.91, p<.05$) across the \textsc{NONE} ($M=.83, SD=1.44$), \textsc{CQ} ($M=.99, SD=.95$), and \textsc{CQGP} ($M=1.53, SD=.82$) conditions. Post-hoc Conover pairwise comparisons revealed a significant difference in \textsc{Pragmatic} quality between the \textsc{CQ} and \textsc{CQGP} conditions ($T=2.74, p<.01$), as well as a significant difference in \textsc{Hedonic} quality between the \textsc{NONE} and \textsc{CQGP} conditions ($T=2.23, p<.05$) as well as between the \textsc{CQ} and \textsc{CQGP} conditions ($T=2.55, p<.05$).
The distribution of overall UEQ-S ratings in all three disambiguation conditions is provided in \Cref{fig:nasa-tlx-ueq-box} (right).

\subsection{Qualitative Feedback}\label{sec:post-exp-comments}

In the post-experience questionnaire, 6 participants (25\%) found clarification questions more helpful, 15 participants (62.5\%) found graphical previews more helpful, while the remaining 3 participants (12.5\%) found both disambiguation techniques equally helpful. When faced with ambiguity, 11 participants (45.8\%) described that they ``relied on trial and error", another 11 participants (45.8\%) described that they ``relied mostly on the graphical preview", while the remaining 2 participants (8.3\%) described that they ``used a mixture of both disambiguation techniques".

A thematic analysis~\cite{guest2012introduction} on post-experience user comments about the interaction process when they completed the task with and without disambiguation support revealed the following trends. Summary of feedback from individual participants grouped by themes and sub-themes are provided in \Cref{tab:quotes}, while the complete set of data collected in the post-experience questionnaire is provided in the Supplemental Materials.

\begin{table*}[h!]
  \caption{Post-experience questionnaire user feedback organized by disambiguation condition and themes (+: positive sentiment, -: negative sentiment, ±: mixed sentiment). Feedback for the \textsc{CQ} and \textsc{CQGP} conditions were merged together.}
  \label{tab:quotes}
  \renewcommand{\arraystretch}{1.5}
  \resizebox{\textwidth}{!}{
  \begin{tabularx}{1.2\textwidth}{| >{\centering\arraybackslash}p{0.17\textwidth}|>{\centering\arraybackslash}p{0.09\textwidth}|p{0.33\textwidth}|X|}
    \hline
\multicolumn{1}{|c|}{\textbf{Theme}}                      & \textbf{Condition}      & \multicolumn{1}{c|}{\textbf{Sub-Theme}}                     & \multicolumn{1}{c|}{\textbf{Summary of User Feedback}}                                                                                                               \\ \hline
\multirow{7.6}{*}{\makecell{Decision-Making\\Approach}}                 & \multirow{3.6}{*}{NONE}   & Iterative   Trial and Error                                 & \leavevmode Relied on issuing many commands to reach desired shape. (P10, -) 
\\ \cline{3-4} 
              &                         & \multirow{1.6}{*}{Progressive   Refinement}                                    & Some refined changes incrementally and started with larger differences. \newline(P20, ±)      \\ \cline{3-4} 
              &                         & Direct   Inquiry                                            & Asked about parameters and their range to guide editing. (P22, +)                                          \\ \cline{2-4} 
              & \multirow{4.6}{*}{CQ\&GP} & Importance of Visual Feedback                             & Previews helped decision-making and increased command specificity. (P24, +)                                    \\ \cline{3-4} 
              &                         & \multirow{1.7}{*}{Mixed Opinions on Clarification}         & Some found clarification questions helpful (P10, P19, +), while others found them confusing because of the technical terms involved. (P4, P13, P16, P20, -)                                                          \\ \cline{3-4} 
              &                         & Improved Confidence on Vague Commands                           & Users felt more confident using vague commands. (P3, +)                                 \\ \cline{3-4} 
              &                         & Reliance on Precise Control                 & Some users still preferred numerical adjustments for precise control. (P7, -)                                \\ \hline
\multirow{6.5}{*}{\makecell{Interaction Pace}}          & \multirow{3.4}{*}{NONE}   & \multirow{2}{*}{Variability   in Speed}                     & Pace ranged among users, from slow and uncertain to quick but imprecise. \newline(P4, P13, ±)                                                                                            \\ \cline{3-4} 
              &                         & \multirow{2}{*}{Autonomy vs. Uncertainty}                     & Some enjoyed the autonomy (P20, +), while others frequently verified details due to uncertainty (P22, -).                           \\ \cline{2-4}
              & \multirow{2.8}{*}{CQ\&GP} & \multirow{2}{=}{Different Effects of CQ and GP on Speed}  & Graphical previews sped up large decisions while clarification questions slowed down small ones. (P9, ±)                                                                                  \\ \cline{3-4} 
              &                         & \multirow{1}{*}{More   Deliberation}                   & Participants reported more thoughtful edits. (P1, P3, +)                                                                                          \\ \hline
\multirow{9}{*}{\makecell{Confidence,\\Error Management,\\and Agency}} & \multirow{4.5}{*}{NONE}   & \multirow{1.6}{=}{Lower Confidence in Command Effectiveness}                 & Users felt unsure whether the assistant could interpret commands accurately. (P11, -)                      \\ \cline{3-4} 
          &                         & Frequent   Need for Corrections                             & Users needed to rephrase commands often to reach desired outcome. (P7, -)                                                                                                                      \\ \cline{3-4} 
          &                         & \multirow{1.6}{*}{Mixed Sense of Control/Agency}                        & Some felt empowered by flexibility but also felt uncertain of the edits made by the system. (P21, ±)                        \\ \cline{2-4} 
          & \multirow{6}{*}{CQ\&GP} & Enhanced   Confidence with Graphical \newline Previews               & Graphical previews provided a way of double-checking and improved confidence. (P16, +)        \\ \cline{3-4} 
          &                         & \multirow{1.6}{=}{Mixed Effects of Clarification Questions} & Some gained clarity (P19, +), while others struggled to phrase commands correctly (P9, -).                                                                                         \\ \cline{3-4} 
          &                         & Reduced   Error Correction             & Edits required less backtracking when disambiguation was present. (P12, +)                                                                       \\ \cline{3-4} 
          &                         & Complexity   and Hesitation when \newline Overwhelmed                & \multirow{1.6}{=}{Some felt burdened by extra choices or unclear guidance. (P23, -)}                  \\ \cline{3-4} 
          &                         & Trade-off   between Control and Guidance                    & Previews increased confidence but sometimes reduced agency. (P21, ±)                  \\ \hline
\end{tabularx}
    }
\end{table*}

\paragraph{Decision-making approach.} When disambiguation was not available, many participants relied on trial and error (P3, P10, P14). They adjusted parameters and observe outcomes and iteratively refined their commands. P9 and P20 also shared that they tend to start with broad, general commands to control the geometry shape and size and then proceed to details such as geometry curvature. Some participants (P8, P13, P22) directly asked for parameter information which reflects how they valued transparency of the internal state of the system.
When disambiguation support was available, many participants (P1, P2, P8) reported that the graphical previews fundamentally changed the way they made decisions by allowing them to rapidly assess the effects of their commands through visual feedback. For clarification questions however, some users found it useful to help them explore the parameter options (P1) and inspired them to ask good questions (P10), while others felt it added complexity or confusion to their workflow (P9, P20). P3 sometimes provided vague instructions because he trusted the system with disambiguation support will guide him towards his intended edit result. Meanwhile, others (P7, P20) preferred to adjust the parameter numerical values directly to make precise edits.

\paragraph{Interaction Pace.} In the \textsc{NONE} condition, some users reported a faster pace in making edits (P1, P13) while others reported being slower due to the increased uncertainty (P2, P4). This reflects a trade-off between the autonomy and fast pace provided by LLM-assisted 3D editing systems against the caution due to uncertainty. P20 felt that the lack of disambiguation promoted a sense of autonomy, while P22 pointed out the need to double-check every detail which led to slow interactions. When disambiguation support was available, participants noted that the graphical previews enabled them to make faster decisions (P1, P9), but claimed that clarification questions tended to slow down the interaction process (P1, P9, P16). Users reported that disambiguation support encouraged them to take more time to review information (P7, P12) and optimize their decisions (P17). However, this can also lead to frustration when the system misinterprets user intent (P23).

\paragraph{User confidence and agency.} In the \textsc{NONE} condition, participants were less confident on whether their vague commands could be interpreted correctly by the system (P1, P4, P11). Participants often needed to revert changes and issue commands again to fix errors (P3, P7, P8, P15, P23). Some participants reported feeling less in control (P4, P22, P24), while others expressed a greater sense of agency because they could freely experiment (P20, P21). When disambiguation support was available, participants felt more confident when they could see graphical previews (P1, P11, P16). Meanwhile, clarification questions improved the certainty of some users (P19), but also introduced complexity and reduced confidence for others (P9). Several participants reported that disambiguation support allowed them to make more precise and deliberate edits, resulting in fewer and easier corrections (P12, P15, P22). Nevertheless, a few users reported feeling disturbed (P20) and less confident (P13, P23), and used more effort to make decisions and edits (P23) with disambiguation support. Some comments revealed that while disambiguation support increased confidence (P17), users' sense of control was sometimes compromised (P21).

\section{Discussion}\label{sec:discussion}

Our study provides new insights into the role of disambiguation techniques in the context of LLM-assisted systems in immersive environments. Unlike prior work which studied disambiguation techniques in dialogue systems \cite{stoyanchev2014towards, alfieri2022intent} or 2D graphical interfaces \cite{li2020multi}, results from quantitative metrics and qualitative user comments reveal distinctive challenges and opportunities on how clarification questions and graphical previews influence the interaction process in LLM-assisted immersive environments. 
% In the following subsections, we present our headline results and propose design implications.

\subsection{User Intent Disambiguation in 3D Space}

In line with the dialogue clarification work by Alfieri et al. \cite{alfieri2022intent}, our qualitative analysis indicates that disambiguation cues can help users refine ambiguous commands. However, our results also indicate that in 3D spaces where spatial reasoning is intrinsic, the nature of ambiguity is different from that in 2D UIs. Users are not only disambiguating textual information or 2D visual information, but are also interpreting spatial cues in real time, as observed in their references to the spatial characteristics of the 3D geometry in their speech commands.

\subsection{Performance Consistency and Efficiency}

A key finding is that clarification questions only (\textsc{CQ} condition) and the combined use of clarification questions and graphical previews (\textsc{CQGP} condition) both significantly reduced task progression variability as evidenced by the MSSD metric, guiding users towards a more stable task progression trajectory compared with the condition without disambiguation support. The maximum closeness score did not differ significantly across conditions, and the \textsc{CQGP} condition required significantly fewer conversation rounds to complete the task, which suggests enhanced interaction efficiency. This finding is novel compared to 2D interfaces, where disambiguation strategies have been shown to improve task performance \cite{li2020multi} and extends earlier work by Horvitz \cite{horvitz1999principles} on mixed-initiative interfaces by demonstrating that in immersive environments, text and visual feedback can guide users towards a smoother task progression trajectory. 

\subsection{Decision-Making and Confidence}

Qualitative feedback revealed that without disambiguation support, users often resorted to trial-and-error. They reported frequent revisions and corrections as they attempted to interpret system feedback. In contrast, when disambiguation support was available, users demonstrated deliberate decision-making. Graphical previews improved their confidence through immediate visual cues about the change to be made. However, some users also noted that too much information can be overwhelming, leading to increased complexity and hesitations. Some also noted that the guided cues occasionally constrained exploratory behavior and reduced the sense of agency.

\subsection{Cognitive Effort}

While no significant differences were found in the overall NASA-TLX rating across the three conditions, post-experience questionnaire comments suggest that some participants took more cognitive effort in decision-making (P1, P3). This indicates that clarification questions and graphical previews can impose additional cognitive demands in immersive settings. Unlike 2D UIs where visual feedback can be processed with relatively low effort, the complexity of spatial environments can cause disambiguation techniques to easily overload the user with too much textual, 2D, and 3D information. This trade-off between confidence and load calls for careful consideration in the design of disambiguation and feedback mechanisms in LLM-assisted systems in immersive environments.

\subsection{Design Recommendations}

Based on our findings, we propose the following design recommendations (DRs) for incorporating disambiguation support in immersive LLM-assisted interactive systems:

\begin{itemize}
    % \item \textbf{DR1: Constrain LLM Actions and Convey in Accessible Language.} According to comments in \Cref{sec:post-exp-comments}, users commonly needed to revert changes to fix errors.
    \item \textbf{DR1: Support Coarse-to-Fine Edit Flows.} The analysis in \Cref{sec:post-exp-comments} indicate users adopt a strategy of naturally shifting from general commands to detailed refinements. Systems should therefore support a nonlinear adjustment process. Systems could allow users to flexibly navigate between LLM-assisted systems for high-level edits and direct manipulation for fine-grained adjustments. Previews of changes and straightforward undo capabilities could reduce the need for extensive error correction.
    \item \textbf{DR2: Provide Adaptive Disambiguation Depth based on Inferred Expertise.} Qualitative feedback  (\Cref{sec:post-exp-comments}) reveal that while novice users may benefit from explicit support, expert users may prefer less interruption. While 15 participants preferred graphical previews, 6 preferred clarification questions, which indicates a different mental model to approach the task. Customizable settings would allow users to adjust the type of disambiguation support to match their expertise and the task complexity.
    \item \textbf{DR3: Balance Guidance with Autonomous Exploration.} Participants' remarks (\Cref{sec:post-exp-comments}) highlight that support mechanisms should not constrain user creativity. Interaction data analysis (\Cref{sec:interaction-data}) further suggests the potential of hybrid disambiguation techniques to support smoother task progression. In addition, the interface should provide fast and responsive feedback, and the disambiguation support should appear as add-ons in the form of nonintrusive side panels which do not disrupt or significantly alter the original interaction flow. This would reduce the disruption to user creativity while keeping the support easily accessible.
    \item \textbf{DR4: Convey LLM Responses in Accessible Language.} In the post-experience questionnaire (\Cref{sec:post-exp-comments}), several users felt confused and overwhelmed by the technical terms involved in the conversation (P4, P13, P16, P20) and stated how the language in clarification questions itself could introduce new ambiguity through internal terminology. This is not a limitation of clarification questions per se, but rather an issue of how the LLM was prompted to generate them. Future systems should instruct the LLM to paraphrase based on the user's vocabulary or include visual cues as in the \textsc{CQGP} condition to lower the vocabulary barrier.
\end{itemize}

\subsection{Limitations and Future Work}

One limitation is the lack of a comparison of the disambiguation techniques in 2D and immersive 3D environments. It is therefore not obvious whether traditional disambiguation techniques can result in improved performance or user experience in immersive environments compared with 2D UIs. This work also hypothesizes that interactions in immersive environments can be represented as a change in latent parameters, which is not necessarily true for some complex scenes. Additionally, we chose to maintain internal validity by having users observe and edit the geometry from the same initial viewpoint, and only slight adjustments of posture were allowed. Therefore, the study does not evaluate disambiguation cases when users have to navigate within the scene, which is open for future work.
Because our research objective was to study how to augment and improve dialogue-based LLM agents in VR, we adopted an additive design (NONE to CQ to CQGP). Therefore, we did not evaluate a graphical-preview-only (GP) condition. It remains an open question for future work to assess whether a pure visual disambiguation interface might outperform the hybrid approach by removing the cognitive friction associated with processing textual output from the system.
We also clarify that this work is different from some other disambiguation works which aim to disambiguate user intent by including additional input modalities such as eye gaze, raycast, or pointing gestures~\cite{bolt1980put, wong2010you, chen2025comparative, lee2024gazepointar}.
In \Cref{sec:performance}, we acknowledge that geometries are not perfectly matched in difficulty and we have attempted to mitigate this limitation through the within-subjects design and order counterbalancing. Additionally, this work does not focus on mitigating LLM hallucinations and more robust safeguards could be further studied.

% \begin{figure}[h!]
%     \centering
%     \includegraphics[width=\linewidth]{figures/living room position-after.png}
%     \includegraphics[width=\linewidth]{figures/living room light-after.png}
%     \caption{Examples of how \textsc{DisambVR} could be applied in VR parameterized editing scenarios to edit object layouts by altering the object positional and rotational coordinates (top) and editing lighting schemes by changing the lighting RGB color parameters and intensity values (bottom).}
%     \label{fig:applications}
% \end{figure}

While our study adopts a controlled geometry editing task, other parameter-driven editing tasks (such as object layout or lighting adjustments) may in principle be represented with the same workflow. Details of whether and how disambiguation techniques transfer to these settings remain as future work.
% we highlight that \textsc{DisambVR} can be easily extended and applied to other parameter-driven VR editing scenarios, such as editing object layouts in a virtual room by updating object positional and rotational coordinates and editing lighting schemes by adjusting RGB and intensity values (\Cref{fig:applications}).

\section{Conclusion}\label{sec:conclusion}

In this paper, we present how disambiguation techniques could be incorporated in LLM-assisted systems in immersive environments. We evaluated the system without disambiguation support, with clarification questions, and with clarification questions and graphical previews in the task where three types of complex geometries are edited using verbal commands to match the target appearance. 
Our findings demonstrate that while combining clarification questions with spatially-anchored graphical previews significantly stabilizes task progression, reduces the number of conversational rounds, and restores user confidence, it introduces a design trade-off. While quantitative metrics (NASA-TLX) indicated no significant increase in overall perceived workload, qualitative feedback revealed that for some users, layering textual, 2D, and 3D feedback imposed additional cognitive demands and risked information overload.

These insights suggest that spatial computing may benefit from a shift away from traditional 2D chatbot paradigms. LLM-assisted editing systems which integrate rich graphical previews with textual guidance have the potential to support smoother task progression and improve user experience in immersive environments.
The design implications derived from our study suggest that future systems could balance real-time visual feedback with effective textual guidance while preserving the user's freedom to explore creative solutions.
We hope this work encourages further exploration of adaptive disambiguation strategies to support richer and more productive experiences in immersive environments.

\section*{Supplemental Materials}
\label{sec:supplemental_materials}

All supplemental materials are available at \textit{$<$anonymized for review$>$}.
These include: (1) Details on the system instructions prompt for \textsc{DisambVR} under different study conditions, (2) Instructions given to participants at the beginning of the study, and (3) Complete sequence of all 24 participants in the user study.

\begin{acks}
% Junlong Chen is supported by the China Scholarship Council and Cambridge Trust. 
The authors thank Jeroen van Ameijde and Provides Ng from the Chinese University of Hong Kong (CUHK) for their contribution to early discussions on this project.
\end{acks}

\bibliographystyle{ACM-Reference-Format}
\bibliography{template}

\appendix

\end{document}